\documentclass[12pt]{iopart}

\usepackage{graphicx}
\usepackage{dcolumn}
\usepackage{bm}
\usepackage{amssymb}
\usepackage{epsf}
\usepackage{color}
\usepackage{booktabs}
\usepackage{cite}
\usepackage[utf8]{inputenc}

\begin{document}

\title[Kinetic Arrest of a Disordered First-Order Phase Transition]{Kinetic Arrest of a Disordered First-Order Phase Transition in Strained $\mathrm{V_2O_3}$ Thin Films}

\author{Sindhunil Barman Roy$^{1,2,3}$}
\address{$^1$ Department of Physics, Ramakrishna Mission Vivekananda Educational and Research Institute, Belur Math, Howrah 711202, India}
\address{$^2$ Department of Physics, Indian Institute of Technology Bombay, Powai, Mumbai 400076, India}
\address{$^3$ UGC-DAE Consortium for Scientific Research, University Campus, Khandwa Road, Indore 452001, India}

\ead{sindhunilbroy@gmail.com}

\date{\today}

\begin{abstract}
We develop a continuum phenomenological theory for the kinetic arrest (KA) of a first-order phase transition (FOPT), taking the thermal and electric field-driven metal-insulator transition (MIT) in epitaxial $\mathrm{V_2O_3}$ thin films as a prototype. By defining a coarse-grained mesoscopic order parameter $\phi(\mathbf{r}, t) \equiv 1 - \theta(\mathbf{r}, t)/\theta_0$ representing the metallic volume fraction coupled to the primary structural monoclinic distortion $\theta$, we formulate a Time-Dependent Ginzburg-Landau (TDGL) framework integrated with a non-local elastic interaction kernel and an Imry-Wortis quenched disorder landscape. We derive an explicit dynamic criterion for the kinetic arrest boundary $T_K(\mathbf{E})$ by analyzing the divergence of the intrinsic order-parameter relaxation time relative to experimental observation timescales. Our framework rigorously decouples the thermodynamic driving force from the kinetic activation barrier. We demonstrate that epitaxial substrate-induced clamping elevates local elastic barriers, pinning domain boundaries and stabilizing an arrested, inhomogeneous mosaic of metallic and insulating phases down to cryogenic temperatures. The simulated transformation kinetics exhibit an Avrami exponent suppression to values $n < 1$, reflecting slow, hindered domain-wall propagation. Furthermore, we examine the effect of an external electric field as an isothermal mechanism that lowers this kinetic barrier, providing a phenomenological description of threshold de-arrest and multi-level resistance states. We critically discuss this electronic barrier-lowering mechanism in comparison with electrothermal Joule heating, establish diagnostic criteria to distinguish them, and contextualize these non-equilibrium dynamics with related phenomena in disordered superconducting films.
\end{abstract}

\vspace{2pc}

\maketitle

\section{Introduction}
First-order phase transition (FOPT) is a ubiquitous phenomenon in condensed matter physics, characterized by discontinuous changes in primary thermodynamic quantities such as entropy, volume, magnetization, or electric polarization as an external intensive variable (temperature, pressure, magnetic field, or electric field) is varied. In solid-state systems, these transitions frequently involve coupled electronic, orbital, magnetic, and structural degrees of freedom. Canonical examples include doped $\mathrm{CeFe_2}$ intermetallics, $\mathrm{Gd_5Ge_4}$, mixed-valence manganites, and Heusler shape-memory alloys, where magnetostructural couplings yield macroscopic functional properties such as colossal magnetoresistance, the giant magnetocaloric effect, and magnetic superelasticity \cite{sbroy2013}. Analogous discontinuous transitions govern field-induced polarization jumps in relaxor ferroelectrics and first-order vortex lattice melting in clean type-II superconductors. A central, unifying characteristic of FOPTs in real crystalline solids is their profound sensitivity to quenched disorder \cite{imry1979}, which smears the transition, introduces a broad distribution of local transition temperatures, and generates persistent phase coexistence and thermodynamic hysteresis (see \cite{sbroy2013} and references therein).

The archetypal correlated Mott insulator $\mathrm{V_2O_3}$ exhibits a canonical first-order Mott metal-insulator transition (MIT) at $T_N \approx 160$~K \cite{yethiraj,leiner}. Thermodynamically driven by a sensitive balance between electron kinetic energy, on-site Coulomb repulsion, and lattice deformation \cite{rozenberg,stewart}, this transition is increasingly recognized as being governed by real-space phase inhomogeneities, mesoscopic phase separation, and spontaneously nanotextured coexistence of correlated metallic and insulating domains \cite{stewart,lupi,frandsen,macleod}. In addition, signatures of magnetic frustration of the MIT have been observed in $\mathrm{V_2O_3}$ \cite{trast}. From an applied perspective, the high sensitivity of the transition to external perturbations makes $\mathrm{V_2O_3}$ a compelling platform for resistive random-access memory (ReRAM), neuromorphic devices, and ultrafast optical switches \cite{kalch}. The electrically driven insulator-to-metal transition (IMT) allows for carrier density modulation exceeding conventional semiconducting channels, placing field-induced Mott gap collapse at the forefront of post-CMOS electronic architectures.

Experimental investigations of epitaxial $\mathrm{V_2O_3}$ thin films have revealed intricate electrothermal history effects, multi-level resistance plateaus, and tunable memristive behavior, typically interpreted via empirical resistor network models or phenomenological nucleation formalisms \cite{kalch2020,de,stolier2014}. The persistent, incomplete transformation observed across broad temperature ranges can be naturally rationalized as a disorder-broadened FOPT interrupted by the dynamic freezing of phase boundaries, a phenomenon termed kinetic arrest (KA) \cite{sbroy2013,roychaddah,chaddah2018}. Kinetic arrest occurs when the intrinsic relaxation time of the order parameter diverges and exceeds the experimental measurement or cooling time window, trapping the system in a non-equilibrium, metastable multi-phase mosaic down to cryogenic temperatures \cite{roychaddah}.

While resistive switching in Mott materials is often strongly influenced by electrothermal Joule heating \cite{stolier2014,valmianski2018}, exploring the isothermal, field-induced modulation of transformation barriers provides fundamental insights into non-equilibrium phase boundaries. In this work, we formulate a continuum phenomenological framework that captures the non-equilibrium transformation kinetics, long-range elastic clamping, and electric-field-induced de-arrest in strained $\mathrm{V_2O_3}$ thin films. We formally decouple the thermodynamic driving force from the kinetic activation barrier, derive an explicit analytical expression for the kinetic arrest boundary $T_K(\mathbf{E})$, incorporate the Imry-Wortis disorder landscape, and account for epitaxial substrate clamping through non-local elasticity. Numerical solutions of the resulting Time-Dependent Ginzburg-Landau (TDGL) equation demonstrate that long-range strain accumulation stabilizes persistent phase coexistence, suppresses the effective Avrami exponent ($n < 1$), and provides a phenomenological foundation for understanding both thermal metastability and threshold switching behavior.

\section{Theoretical Formalism}

\subsection{Physical Definition of the Order Parameter and Multi-Order Coupling}
In $\mathrm{V_2O_3}$, the first-order MIT is multiferroic in character, coupling three distinct physical sectors:
\begin{enumerate}
    \item {\it Lattice Distortion ($\theta$):} A symmetry-breaking structural transformation from the high-temperature rhombohedral ($R\bar{3}c$) phase to the low-temperature monoclinic ($I2/a$) phase.
    \item {\it Electronic Localization ($n_d$):} The opening of a Mott-Hubbard correlation gap ($U \gg W$) and the collapse of the electronic density of states at the Fermi level.
    \item {\it Magnetic Ordering ($M$):} The transition from a high-temperature paramagnetic state to a low-temperature antiferromagnetic insulator.
\end{enumerate}
Because these distinct order parameters are strongly cross-coupled at the microscopic level through terms of the form $-\gamma_1 \theta^2 M^2 - \gamma_2 \theta n_d$, they jump synchronously across the first-order boundary. To proceed further, we take the structural monoclinic distortion $\theta$ as the primary master order parameter that coordinates electronic localization and magnetic order.

To model the kinetics across mesoscopic length scales ($\sim$ 10 - 100 nm) where elastic clamping and domain boundaries dominate, we define a normalized, coarse-grained scalar field:
\begin{equation}
    \phi(\mathbf{r}, t) \equiv 1 - \frac{\theta(\mathbf{r}, t)}{\theta_0} \in [0, 1],
    \label{eq:order_param_def}
\end{equation}
where $\theta_0$ is the equilibrium monoclinic strain distortion of the fully insulating ground state. In this normalized representation:
\begin{itemize}
    \item $\phi = 0$ ($\theta = \theta_0$) represents the low-temperature monoclinic, antiferromagnetic insulating (AFI) phase.
    \item $\phi = 1$ ($\theta = 0$) represents the high-temperature rhombohedral, paramagnetic metallic (PM) phase.
    \item $0 < \phi < 1$ describes the local volume fraction of the metallic phase within a mesoscopic coarse-graining volume element $V_{cg} \sim \xi_0^3$.
\end{itemize}

\subsection{Thermodynamic Potential and Analytic Landscape Analysis}
To capture the discontinuous nature of the transition where heating drives the system from the low-temperature insulating phase($\phi = 0$) to the high-temperature metallic phase ($\phi > 0$), we express the local bulk free-energy density as a standard 2-4-6 Landau polynomial \cite{chaiklub}:
\begin{equation}
    f_0(\phi, T) = \frac{a(T)}{2}\phi^2 - \frac{b}{4}\phi^4 + \frac{c}{6}\phi^6,
    \label{eq:landau_f0}
\end{equation}
where $b > 0$ and $c > 0$ are temperature-independent phenomenological coefficients ensuring a first-order topology and high-amplitude stability, respectively. To correctly reflect that thermal energy stabilizes the metallic phase, the temperature dependence enters via the quadratic coefficient:
\begin{equation}
    a(T) = a_0(T_0 - T),
    \label{eq:a_T}
\end{equation}
with $a_0 > 0$, where $T_0$ represents the upper spinodal threshold on heating.

\subsubsection{Critical Temperatures and Spinodals}
The stationary extrema of $f_0(\phi, T)$ are governed by:
\begin{equation}
    \frac{\partial f_0}{\partial \phi} = \phi \left( a(T) - b\phi^2 + c\phi^4 \right) = 0.
    \label{eq:df_dphi}
\end{equation}
Eqn. 4 admits the trivial root $\phi_1 = 0$ (insulating state) and non-trivial roots given by:
\begin{equation}
    \phi_{\pm}^2(T) = \frac{b \pm \sqrt{b^2 - 4a(T)c}}{2c}.
    \label{eq:phi_pm}
\end{equation}
These roots analytically define three critical temperature benchmarks:
\begin{itemize}
    \item {\it Upper Spinodal ($T_0$ / Heating Spinodal):} For $T > T_0$, $a(T) < 0$, rendering the insulating state ($\phi = 0$) an unstable local maximum and leaving the metallic phase ($\phi > 0$) as the unconditionally stable thermodynamic state.
    \item {\it Critical Temperature ($T_c$):} The thermodynamic equilibrium transition occurs when the free energy of the metallic minimum matches that of the insulating ground state, $f_0(\phi_+, T_c) = f_0(0, T_c) = 0$:
    \begin{equation}
        a(T_c) = \frac{3b^2}{16c} \ implies \ T_c = T_0 - \frac{3b^2}{16a_0 c}.
        \label{eq:T_c}
    \end{equation}
    At $T = T_c$, the equilibrium order parameter jumps discontinuously to:
    \begin{equation}
        \phi_c = \sqrt{\frac{3b}{4c}}.
        \label{eq:phi_c}
    \end{equation}
    \item {\it Lower Spinodal ($T^*$ / Cooling Spinodal):} Setting the discriminant in Eq.5 to zero ($b^2 - 4a(T)c = 0$) defines the lower limit of metastability (supercooling limit) for the metallic phase:
    \begin{equation}
        T^* = T_0 - \frac{b^2}{4a_0 c}.
        \label{eq:T_star}
    \end{equation}
    For $T < T^*$, the non-trivial roots cease to be real-valued, meaning the metallic local minimum completely vanishes and the insulating phase ($\phi = 0$) becomes the only stable extremum.
\end{itemize}

\subsubsection{Decoupling Thermodynamic Driving Force ($\Delta f$) and Kinetic Barrier ($\Delta f^{\ddagger}$)}
A central physical requirement is to distinguish the thermodynamic impetus to transform from the dynamic rate of transformation:
\begin{itemize}
    \item {\it Thermodynamic Driving Force ($\Delta f$):} The free-energy density difference between the competing phases:
    \begin{equation}
        \Delta f(T) = f_0(0, T) - f_0(\phi_+, T) = -\left( \frac{a(T)}{2}\phi_+^2 - \frac{b}{4}\phi_+^4 + \frac{c}{6}\phi_+^6 \right).
        \label{eq:driving_force}
    \end{equation}
    For $T < T_c$, $\Delta f(T) < 0$, making the insulating phase the global thermodynamic ground state. As $T$ decreases well below $T_c$, $|\Delta f(T)|$ increases monotonically, providing an increasingly large thermodynamic driving force toward full transformation.
    \item {\it Intrinsic Kinetic Barrier ($\Delta f^{\ddagger}$):} The intermediate energy barrier between the two minima is set by the local maximum at $\phi_-$:
    \begin{equation}
        \Delta f^{\ddagger}(T) = f_0(\phi_-, T) - f_0(\phi_+, T) = \frac{\left[ b^2 - 4 a_0(T_0 - T) c \right]^{3/2}}{24 c^2}.
        \label{eq:barrier_explicit}
    \end{equation}
\end{itemize}
While cooling below $T_c$ enhances the thermodynamic driving force $|\Delta f(T)|$, the dynamic transformation path is governed by the ratio $(\Delta f^{\ddagger}(T) + U_{el})/(k_B T)$. Here, $U_{el}$ is the elastic activation barrier. When $k_B T \ll \Delta f^{\ddagger} + U_{el}$, thermal fluctuations become insufficient to activate the phase boundary, resulting in dynamic freezing despite a large thermodynamic impetus.

\subsection{Quenched Disorder: Imry-Wortis Formulation}
In reality thin films of V$_2$O$_3$, spatial variations in oxygen vacancy concentration, local stoichiometry, and defect networks induce mesoscopic variations in the transition temperature. Following the Imry-Wortis framework for disordered first-order phase transitions \cite{imry1979}, we introduce a spatially correlated local critical temperature field, $T_c(\mathbf{r}) = T_c^0 + \delta T_c(\mathbf{r})$, into the quadratic term:
\begin{equation}
    a(\mathbf{r}, T) = a_0 \left[ T_0 - T - \delta T_c(\mathbf{r}) \right].
    \label{eq:a_r_T}
\end{equation}
The spatial fluctuations $\delta T_c(\mathbf{r})$ are modeled as a Gaussian random field with zero mean and two-point spatial correlation:
\begin{equation}
    \langle \delta T_c(\mathbf{r}) \rangle = 0, \quad \langle \delta T_c(\mathbf{r}) \delta T_c(\mathbf{r}') \rangle = \sigma_T^2 \exp\left( -\frac{|\mathbf{r} - \mathbf{r}'|^2}{2 \xi_{dis}^2} \right),
    \label{eq:disorder_corr}
\end{equation}
where $\sigma_T$ is the disorder strength parameterizing the local $T_c$ distribution width, and $\xi_{dis}$ is the spatial correlation length corresponding to typical grain or coherent defect domain sizes. Local regions with $\delta T_c(\mathbf{r}) > 0$ serve as deterministic ``hot spots'' that lower the local nucleation barrier, promoting preferential nucleation during the transition.

\subsection{Thin-Film Elastic Energy and Substrate Clamping Boundary Conditions}
The structural transformation in $\mathrm{V_2O_3}$ involves an anisotropic, stress-free transformation strain (Bain strain) tensor:
\begin{equation}
    \epsilon_{ij}^0 = \mathrm{diag}(\epsilon_{xx}^0, \, \epsilon_{yy}^0, \, \epsilon_{zz}^0),
    \label{eq:bain_tensor}
\end{equation}
accompanied by a unit cell volume expansion $\Delta V/V = \mathrm{Tr}(\epsilon_{ij}^0) \approx +1.4\%$. For an epitaxial thin film of thickness $h_f$ grown on a thick, rigid substrate of thickness $h_s$ ($h_f \ll h_s$), the total strain $\epsilon_{ij}(\mathbf{r})$ is decomposed into elastic strain $e_{ij}(\mathbf{r})$ and transformation self-strain:
\begin{equation}
    \epsilon_{ij}(\mathbf{r}) = e_{ij}(\mathbf{r}) + \epsilon_{ij}^0 \, \phi(\mathbf{r}).
    \label{eq:strain_decomp}
\end{equation}

The mechanical boundary conditions are established as:
\begin{enumerate}
    \item {\it In-Plane Substrate Clamping ($xy$-plane):} Rigid coherence with the substrate requires:
    \begin{equation}
        \bar{\epsilon}_{xx} = \bar{\epsilon}_{yy} = \bar{\epsilon}_{xy} = 0,
        \label{eq:inplane_clamp}
    \end{equation}
    where overbars denote macroscopic spatial averages across the film plane.
    \item {\it Out-of-Plane Free Surface Relaxation ($z$-axis):} Traction-free conditions at the film surface ($\sigma_{iz}(z = h_f) = 0$) permit stress relief along the normal through Poisson relaxation:
    \begin{equation}
        e_{zz}(\mathbf{r}) = -\frac{\nu}{1 - \nu} \left[ e_{xx}(\mathbf{r}) + e_{yy}(\mathbf{r}) \right] - \epsilon_{zz}^0 \, \phi(\mathbf{r}),
        \label{eq:poisson_relax}
    \end{equation}
    where $\nu \approx 0.30$ is Poisson's ratio for $\mathrm{V_2O_3}$.
\end{enumerate}

Enforcing mechanical equilibrium ($\nabla_j \sigma_{ij} = 0$) allows the displacement fields to be solved in Fourier space via the Khachaturyan framework \cite{khachaturyan}. Integrating out the elastic fields yields the non-local elastic energy functional:
\begin{equation}
    F_{\mathrm{el}}[\phi] = \frac{1}{2} \int \frac{d^2\mathbf{k}}{(2\pi)^2} B(\mathbf{n}) |\tilde{\phi}(\mathbf{k})|^2,
    \label{eq:F_el}
\end{equation}
where $\mathbf{n} = \mathbf{k}/|\mathbf{k}|$ is the unit wavevector, $\tilde{\phi}(\mathbf{k})$ is the Fourier transform of $\phi(\mathbf{r}) - \bar{\phi}$, and $B(\mathbf{n}) = C_{ijkl}\epsilon_{ij}^0\epsilon_{kl}^0 - n_i \sigma_{ij}^0 \Omega_{jk}(\mathbf{n}) \sigma_{kl}^0 n_l$ is the orientation-dependent elastic kernel with acoustic Green's tensor $\Omega_{jk}^{-1}(\mathbf{n}) = C_{jmkn} n_m n_n$. Because $B(\mathbf{n}) > 0$ along misaligned directions, expanding domains minimize elastic penalties by adopting anisotropic, plate-like morphologies along elastically soft directions. Overlapping strain fields between neighboring domains generate self-stress, creating the elastic activation barrier $U_{el}(\mathbf{r})$.

\subsection{Time-Dependent Dynamics and Kinetic Arrest Criterion}
The non-equilibrium spatio-temporal evolution of $\phi(\mathbf{r}, t)$ is governed by the Time-Dependent Ginzburg-Landau (TDGL) equation:
\begin{equation}
    \frac{\partial \phi(\mathbf{r}, t)}{\partial t} = -\Gamma(T, \mathbf{E}) \frac{\delta F[\phi]}{\delta \phi(\mathbf{r}, t)} + \zeta(\mathbf{r}, t),
    \label{eq:TDGL}
\end{equation}
where the total free-energy functional is:
\begin{equation}
    F[\phi] = \int d^2\mathbf{r} \left[ f_0(\phi, \mathbf{r}, T) + \frac{\kappa}{2}|\nabla \phi|^2 \right] + F_{\mathrm{el}}[\phi] - \int d^2\mathbf{r} \, \mathcal{E}_{ext} \phi.
    \label{eq:F_total}
\end{equation}
Here, $\kappa$ parameterizes the interfacial gradient energy, $\mathcal{E}_{ext} = \mathbf{p}\cdot\mathbf{E}$ accounts for the linear coupling of the external electric field $\mathbf{E}$ to the local dipole/polarization difference $\mathbf{p}$, and $\zeta(\mathbf{r}, t)$ is a conservative Gaussian white noise satisfying the Fluctuation-Dissipation Theorem:
\begin{equation}
    \langle \zeta(\mathbf{r}, t) \zeta(\mathbf{r}', t') \rangle = 2 k_B T \Gamma(T, \mathbf{E}) \delta(\mathbf{r} - \mathbf{r}') \delta(t - t').
    \label{eq:FDT}
\end{equation}

The kinetic coefficient $\Gamma(T, \mathbf{E})$ reflects thermally activated barrier crossing:
\begin{equation}
    \Gamma(T, \mathbf{E}) = \Gamma_0 \exp\left( -\frac{U_{eff}(\mathbf{r}, \mathbf{E})}{k_B T} \right),
    \label{eq:gamma_arrhenius}
\end{equation}
where $U_{eff}(\mathbf{r}, \mathbf{E}) = \Delta f^{\ddagger}(T) + U_{el}(\mathbf{r}) - \alpha |\mathbf{E}|$, with $\alpha$ representing the effective field-lowering coefficient.

\subsubsection{Dynamic Derivation of the Kinetic Arrest Boundary $T_K(\mathbf{E})$}
The characteristic relaxation time $\tau$ governing the local order-parameter dynamics near a metastable state is:
\begin{equation}
    \tau(\mathbf{r}; T, \mathbf{E}) \sim \frac{1}{\Gamma(T, \mathbf{E}) \left| \frac{\partial^2 f}{\partial \phi^2} \right|} \approx \tau_0 \exp\left( \frac{U_{eff}(\mathbf{r}, \mathbf{E})}{k_B T} \right),
    \label{eq:tau_def}
\end{equation}
where $\tau_0 \sim (\Gamma_0 |f''|)^{-1}$ is the microscopic attempt time ($\sim 10^{-12}$~s). Kinetic arrest represents a dynamic crossover where the intrinsic relaxation time $\tau(T, \mathbf{E})$ exceeds the characteristic experimental observation time window $t_{exp}$ (or cooling rate scale $t_{cool} = \Delta T / |\dot{T}|$):
\begin{equation}
    \tau(T_K, \mathbf{E}) \approx t_{exp}.
    \label{eq:arrest_cond}
\end{equation}
Equating Eq.22 to $t_{exp}$ yields the analytical expression for the kinetic arrest boundary $T_K(\mathbf{E})$:
\begin{equation}
    T_K(\mathbf{E}) = \frac{\Delta f^{\ddagger}(T_K) + \langle U_{el} \rangle - \alpha |\mathbf{E}|}{k_B \ln\left( t_{exp} / \tau_0 \right)}.
    \label{eq:T_K_explicit}
\end{equation}
For $T < T_K(0)$, the interface velocity $v \propto \Gamma (-\delta F/\delta \phi) \to 0$ on experimental timescales, freezing the domain pattern into an arrested state. Applying an external electric field ($\mathbf{E} > 0$) lowers $U_{eff}$, depressing $T_K(\mathbf{E})$ and driving field-induced de-arrest (threshold switching).

\section{Numerical Methodology and Implementation}

\subsection{Discretization and Non-Dimensionalization}
The TDGL equation is non-dimensionalized using the characteristic energy density $f_0^* = b^2/(4c)$, length scale $\xi_0 = \sqrt{\kappa/f_0^*}$, and relaxation time $\tau_0 = (\Gamma_0 f_0^*)^{-1}$:
\begin{equation}
    \mathbf{r}^* = \frac{\mathbf{r}}{\xi_0}, \quad t^* = \frac{t}{\tau_0}, \quad f^* = \frac{f}{f_0^*}, \quad F^* = \frac{F}{f_0^* \xi_0^2}.
    \label{eq:nondim}
\end{equation}
For $\mathrm{V_2O_3}$, physical estimates correspond to $\xi_0 \approx 2.5$~nm, $\tau_0 \approx 1.2$~ps, and $f_0^* \approx 4.8 \times 10^7$~J/m$^3$. In dimensionless form:
\begin{equation}
    \frac{\partial \phi}{\partial t^*} = -\tilde{\Gamma}(T, \mathbf{E}) \left[ a^*(T, \mathbf{r}^*)\phi - \phi^3 + \phi^5 - \nabla^{*2}\phi + \mu_{el}^*(\mathbf{r}^*) - \mathcal{E}_{ext}^* \right] + \zeta^*(\mathbf{r}^*, t^*).
    \label{eq:dimless_TDGL}
\end{equation}

\begin{table}[t]
\caption{\label{tab:params}Numerical and physical parameters used in the TDGL simulations for epitaxial $\mathrm{V_2O_3}$ thin films.}
\begin{indented}
\item[]\begin{tabular}{@{}lllll}
\br
Parameter & Symbol & Dimensionless & Physical Estimate & Description \\
\mr
Lattice Size & $N_x \times N_y$ & $256 \times 256$ & $512 nm \times 512 nm$ & 2D film cross-section \\
Grid Spacing & $\Delta x^*, \Delta y^*$ & $0.80$ & $2.0 nm$ & Mesh increment \\
Time Step & $\Delta t^*$ & $0.01$ & $12  fs $ & Time marching increment \\
Domain Wall Width & $\xi_0$ & $1.0$ & $2.5  nm$ & Correlation length \\
Mobility & $\Gamma_0$ & $1.0$ & $1.7 \times 10^{-5}  m^3 J \cdot s)$ & Unclamped interface mobility \\
Interfacial Energy & $\kappa^*$ & $1.0$ & $3.0 \times 10^{-10} J/m$ & Gradient coefficient \\
Elastic Modulus & $B_0^*$ & $1.20$ & $5.7 \times 10^7  J m ^3$ & Elastic kernel amplitude \\
Transformation Strain & $\epsilon_0$ & --- & $1.4\%$ & Unconstrained Bain strain \\
Disorder Amplitude & $\sigma_T^*$ & $0.15$ & $6.5 K $ & Width of local $T_c$ fluctuations \\
Disorder Correlation & $\xi_{dis}^*$ & $6.0$ & $15.0  nm$ & Spatial defect coherence \\
Field Coupling & $\alpha$ & $0.08$ & $3.75 e \cdot nm$ & Effective dipole parameter \\
Ensemble Realizations & $N_{runs}$ & $50$ & $50$ & Disorder configurations \\
\br
\end{tabular}
\end{indented}
\end{table}

The simulation domain is discretized on a uniform $256 \times 256$ lattice with periodic boundary conditions. The grid spacing $\Delta x^* = \Delta y^* = 0.80$ corresponds to a physical area of $512 nm \times 512  nm$, ensuring the domain interface width ($\approx 2 - 3\xi_0$) is resolved across multiple grid nodes to prevent spurious lattice pinning. The Laplacian $\nabla^{*2}\phi$ is computed via a 5-point finite-difference stencil, and the non-local elastic stress $\mu_{el}^* = \delta F_{el}^*/\delta \phi$ is evaluated in reciprocal space using 2D Fast Fourier Transforms (FFT).

Temporal marching is executed using the stochastic Euler-Maruyama algorithm:
\begin{equation}
    \phi_{i,j}(t^* + \Delta t^*) = \phi_{i,j}(t^*) - \Delta t^* \tilde{\Gamma}_{i,j} \left[ \frac{\delta F^*}{\delta \phi} \right]_{i,j} + \sqrt{\frac{2 k_B T^* \tilde{\Gamma}_{i,j} \Delta t^*}{(\Delta x^*)^2}} \, \mathcal{N}_{i,j}(0, 1),
    \label{eq:euler_maruyama}
\end{equation}
where $\mathcal{N}_{i,j}(0, 1)$ represents standard normal random variables. Numerical stability for the explicit parabolic scheme requires satisfying the Courant-Friedrichs-Lewy (CFL) condition:
\begin{equation}
    \Delta t^* < \frac{(\Delta x^*)^2}{4 \tilde{\Gamma}_{\max}} \approx \frac{(0.8)^2}{4 \times 1.0} = 0.16.
    \label{eq:CFL}
\end{equation}
Our choice of $\Delta t^* = 0.01$ is well below this stability threshold. Simulation parameters are summarized in Table~\ref{tab:params}.

\section{Results and Discussion}

\subsection{Microscopic Nucleation at Disorder Hot Spots}
\begin{figure}[t]
\centering
\includegraphics[width=0.85\linewidth]{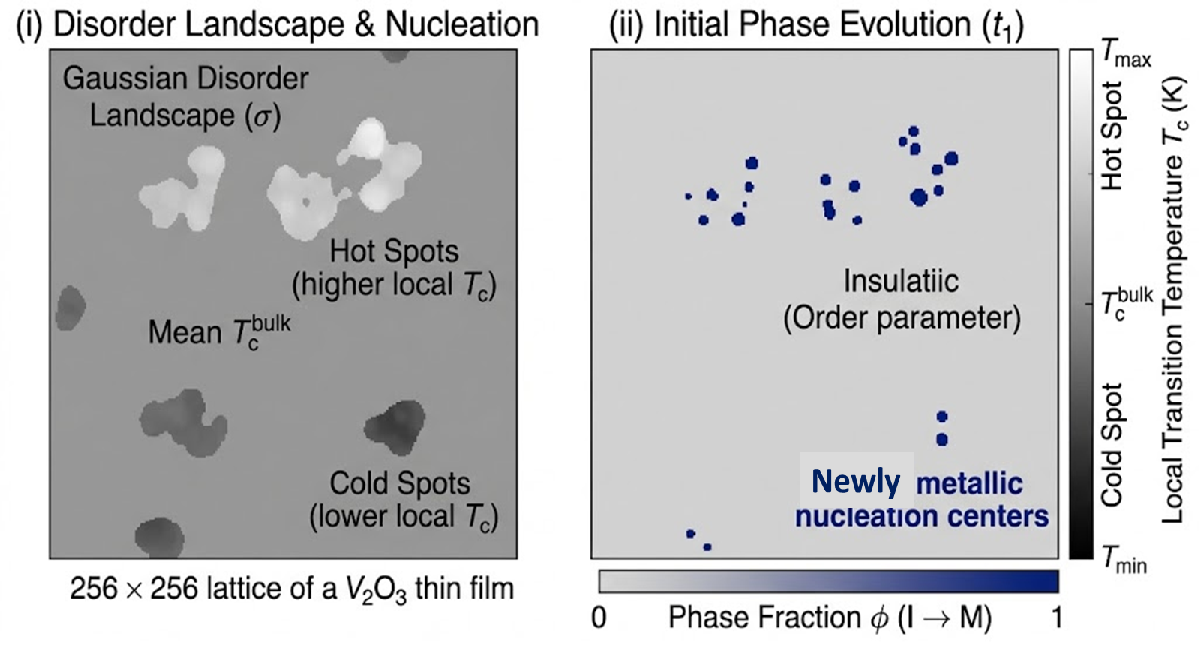}
\caption{Quenched disorder and localized nucleation in epitaxial $\mathrm{V_2O_3}$: (i) Spatial distribution of the Gaussian disorder landscape $\delta T_c(\mathbf{r})$, where lighter tones denote local ``hot spots'' with elevated transition temperatures. (ii) Spatial map of the order parameter $\phi(\mathbf{r})$ at early time $t_1$, showing that metallic nucleation droplets (dark blue) emerge deterministically at these disorder-induced hot spots.}
\label{fig:Fig1}
\end{figure}

Figure~\ref{fig:Fig1} illustrates the microscopic onset of the field-induced transition on a $256 \times 256$ grid. The system is initialized in the insulating state ($\phi \approx 0$) at $T < T_c$, with a sub-threshold bias field applied to probe disorder effects. Panel (i) depicts the Imry-Wortis disorder landscape $\delta T_c(\mathbf{r})$, where intensity variations reflect local shifts in the transition temperature. Lighter regions denote local ``hot spots'' with higher $T_c$, presenting lower thermodynamic nucleation barriers.

Panel (ii) shows the corresponding order-parameter configuration $\phi(\mathbf{r})$ at time $t_1$. A direct spatial comparison demonstrates that initial metallic nuclei ($\phi \approx 1$, dark blue) form predominantly at these hot spots. The spatial profile of quenched disorder dictates the heterogeneous nucleation pattern, illustrating how spatial randomness smears the sharp first-order transition into an inhomogeneous, percolative process.

\subsection{Elastic Clamping and Nanotextured Domain Pinning}
\begin{figure}[t]
\centering
\includegraphics[width=0.85\linewidth]{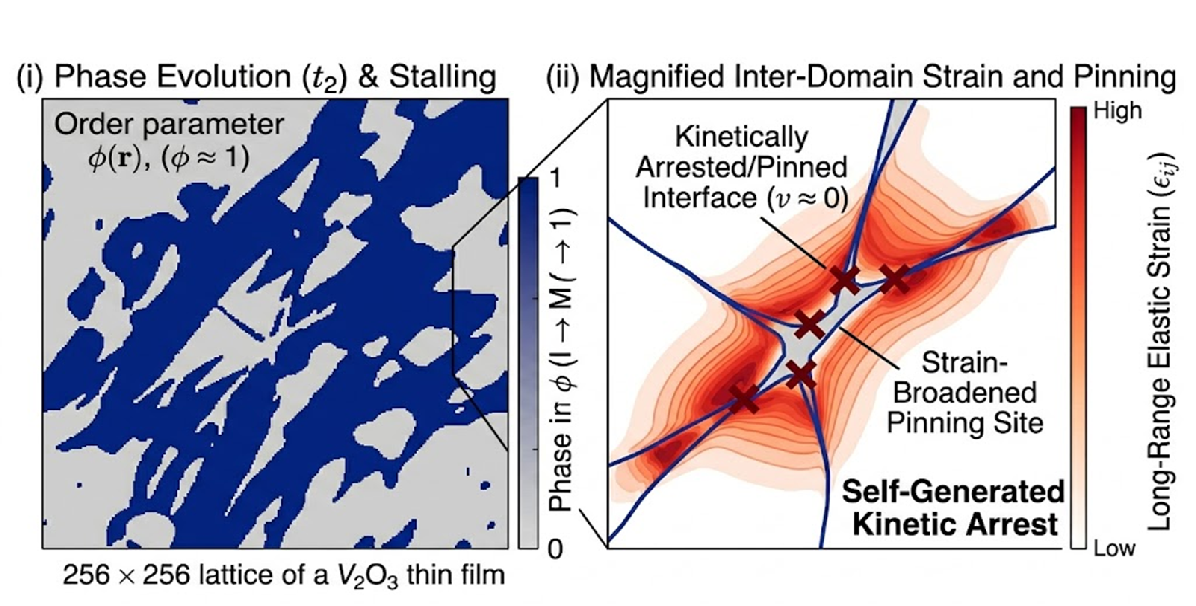}
\caption{Elastic clamping and phase boundary pinning: (i) Spatial phase distribution at later time $t_2$, showing anisotropic, nanotextured metallic domains (dark blue) coexisting with insulating regions (light grey). (ii) Magnified view of inter-domain channels, where overlapping elastic strain fields (red/orange contours) generate an insurmountable activation barrier ($U_{eff}$), pinning domain interfaces (red crosses) and causing kinetic arrest.}
\label{fig:Fig2}
\end{figure}

As metallic domains expand, long-range elastic strain fields fundamentally alter their growth morphology. Because of the lattice mismatch between rhombohedral and monoclinic phases, domain growth generates an anisotropic strain field in the surrounding matrix. Figure~\ref{fig:Fig2}(i) displays the domain morphology at a later time $t_2$. Rather than forming isotropic, circular droplets, the metallic domains develop elongated, plate-like geometries with jagged boundaries, reflecting elastic energy minimization along soft crystallographic directions.

Crucially, the transformation stalls at an incomplete phase fraction ($X \approx 0.7$). Figure~\ref{fig:Fig2}(ii) presents a magnified view of the inter-domain boundaries. As neighboring metallic domains approach, their overlapping strain fields (red/orange contours) significantly increase the local activation barrier $U_{eff}$. This interface pinning (indicated by red crosses) drives the domain boundary velocity toward zero, arresting the system in a non-equilibrium state of persistent phase coexistence.

\subsection{Statistical Master Solutions and Memristive Switching}
\begin{figure}[t]
\centering
\includegraphics[width=0.85\linewidth]{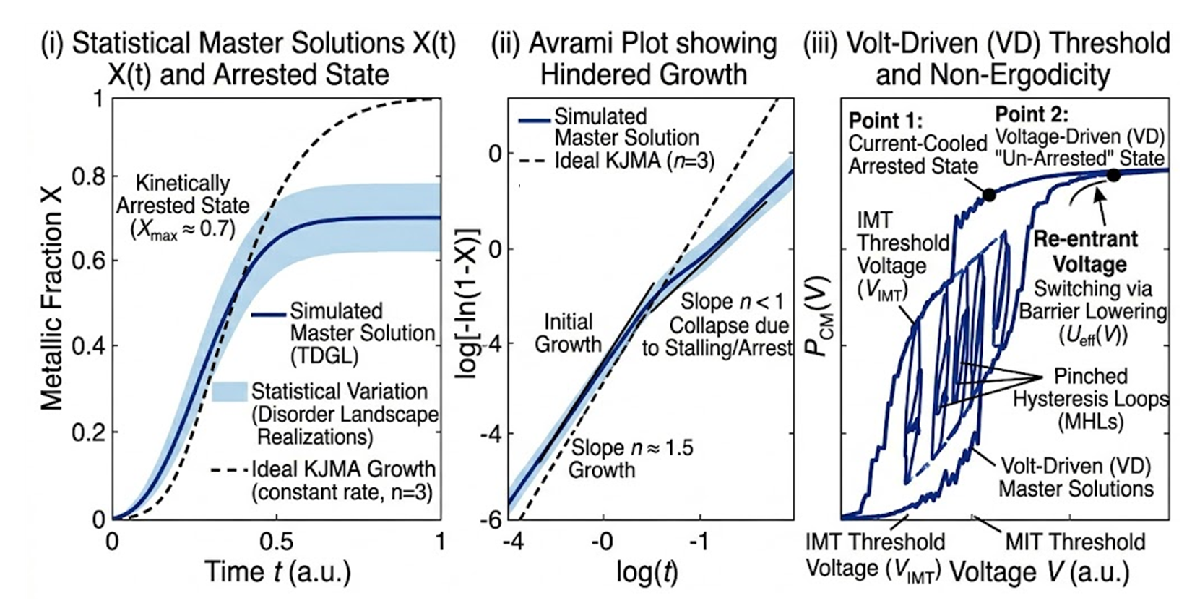}
\caption{Transformation kinetics and memristive switching: (i) Global transformed fraction $X(t)$ comparing ideal 3D KJMA growth ($n=3$, dashed black) with the simulated TDGL master solution (solid blue) stalling at $X_{stall} \approx 0.7$. The shaded band shows statistical variation across 50 disorder realizations. (ii) Avrami plot ($\ln[-\ln(1-X)]$ vs. $\ln t$) showing an initial growth stage ($n \approx 1.5 - 2$) followed by dynamic slope collapse ($n_{eff} < 1$) in the arrested regime. (iii) Voltage-driven minor hysteresis loops (MHLs) in the $I$-$V$ plane showing reproducible de-arrest across threshold voltages ($V_{IMT}$ and $V_{MIT}$) and multi-level resistance states.}
\label{fig:Fig3}
\end{figure}

The macroscopic transformed metallic fraction is evaluated as:
\begin{equation}
    X(t) = \frac{1}{L^2} \int \phi(\mathbf{r}, t) \, d^2\mathbf{r}.
    \label{eq:X_t}
\end{equation}
In standard Kolmogorov-Johnson-Mehl-Avrami (KJMA) theory \cite{kiana}, transformation kinetics follow:
\begin{equation}
    X(t) = 1 - \exp\left( -K t^n \right),
    \label{eq:KJMA}
\end{equation}
where $K$ is a kinetic rate constant and $n$ is the Avrami exponent reflecting growth dimensionality.

Figure~\ref{fig:Fig3}(i) compares the ideal 3D KJMA profile ($n=3$) against our ensemble-averaged TDGL master solution (averaged over 50 disorder realizations). The simulated curve shows a characteristic sigmoidal increase that stalls well below full transformation at $X_{stall} \approx 0.7$, providing macroscopic statistical evidence of kinetic arrest. The shaded envelope reflects statistical variations across different disorder realizations.

Figure~\ref{fig:Fig3}(ii) presents the corresponding Avrami analysis ($\ln[-\ln(1-X)]$ versus $\ln t$). While an unhindered transition maintains a steep, constant slope, the simulated curve reveals two distinct kinetic regimes: an initial growth stage with effective exponent $n \approx 1.5 - 2$ (reflecting site-saturated 2D growth from disorder hot spots), followed by a pronounced slope collapse where the instantaneous effective exponent $n_{eff} = \frac{d\ln[-\ln(1-X)]}{d\ln t}$ drops well below unity ($n < 1$). This sub-unitary exponent is the direct kinetic hallmark of self-generated elastic clamping and domain-wall pinning.

Furthermore, Figure~\ref{fig:Fig3}(iii) illustrates the electrical de-arrest dynamics, plotting the resulting pinched minor hysteresis loops (MHLs) in the current-voltage plane. Applying an external bias lowers the local barriers ($U_{eff} - \alpha|\mathbf{E}|$), dynamically de-arresting pinned boundaries and enabling progressive transformation across threshold voltages ($V_{IMT}$ and $V_{MIT}$). This field-tunable de-arrest accounts for the reproducible multi-level resistance states observed experimentally in $\mathrm{V_2O_3}$ memristive devices.

\begin{figure}[t]
\centering
\includegraphics[width=0.75\linewidth]{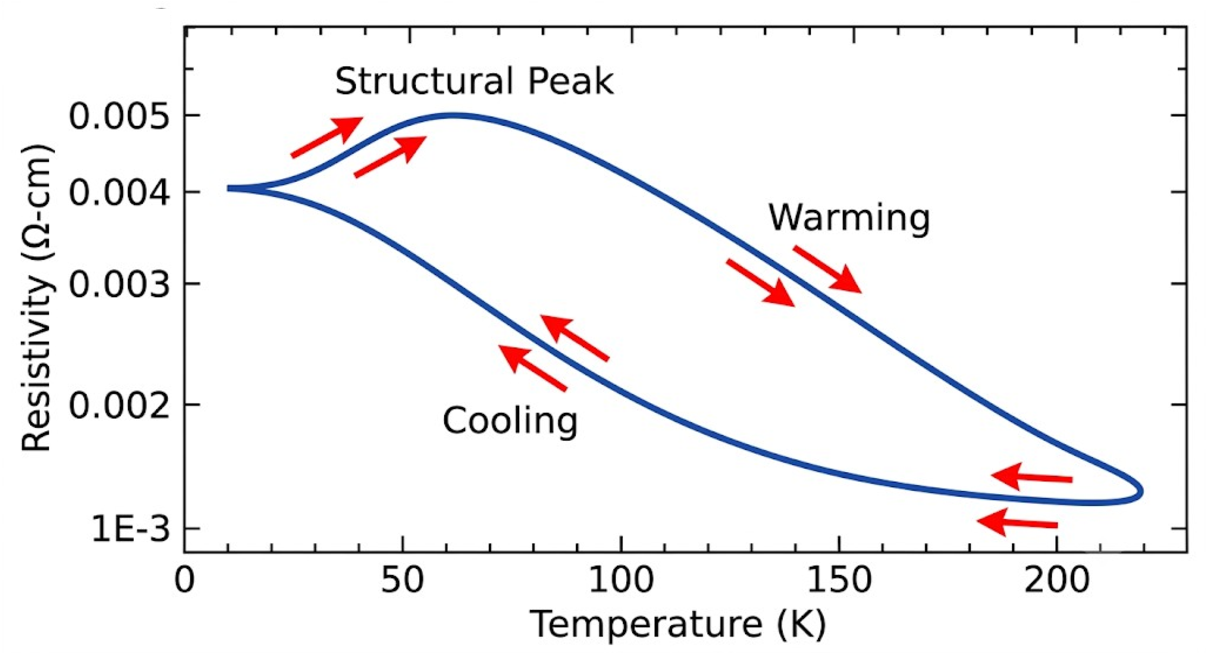}
\caption{Temperature dependence of electrical resistivity $\rho(T)$ across the first-order metal-insulator transition in strained epitaxial $\mathrm{V_2O_3}$ thin films. The cooling branch exhibits dynamic kinetic arrest below $T_K \approx 50$~K, freezing into an incomplete metallic volume fraction ($X_{\mathrm{stall}} \approx 0.60$) and yielding a residual resistivity plateau ($\rho \approx 4.05 \times 10^{-3}~\Omega\cdot\mathrm{cm}$) rather than an exponential divergence. The warming branch reproduces the delayed thermal de-arrest and the characteristic structural relaxation maximum at $T \approx 55$~K before closing the hysteresis loop near $185$~K.}
\label{fig:Fig4}
\end{figure}

\subsection{Temperature Dependence of Resistivity across the First-Order Transition}
To validate our mesoscopic model against macroscopic transport measurements, we derive the explicit temperature-dependent resistivity $\rho(T)$ from $300$~K down to $5$~K by integrating the order-parameter dynamics with effective medium theory. In thin films clamped to rigid sapphire substrates, the material forms an inhomogeneous composite consisting of correlated rhombohedral metallic domains and monoclinic insulating regions. The effective conductivity $\sigma_{\mathrm{eff}}(T) = 1/\rho(T)$ is governed by the symmetric two-dimensional Bruggeman Effective Medium Approximation (EMA):
\begin{equation}
    X(T) \, \frac{\sigma_m(T) - \sigma_{\mathrm{eff}}(T)}{\sigma_m(T) + \sigma_{\mathrm{eff}}(T)} + [1 - X(T)] \, \frac{\sigma_i(T) - \sigma_{\mathrm{eff}}(T)}{\sigma_i(T) + \sigma_{\mathrm{eff}}(T)} = 0,
    \label{eq:bruggeman_rhoT}
\end{equation}
where $\sigma_m(T) = (\rho_{m0} + A_m T)^{-1}$ represents the metallic conductivity ($\rho_{m0} \approx 1.2 \times 10^{-3}~\Omega\cdot\mathrm{cm}$), and $\sigma_i(T)$ is the conductivity of the strained insulating phase. Solving Eq.31 yields the closed-form expression:
\begin{equation}
    \rho(T) = \frac{2}{\mathcal{A}(T) + \sqrt{\mathcal{A}^2(T) + 4 \sigma_m(T) \sigma_i(T)}},
    \label{eq:rho_explicit}
\end{equation}
where $\mathcal{A}(T) \equiv [2X(T) - 1][\sigma_i(T) - \sigma_m(T)]$.

The evolution of the metallic volume fraction $X(T)$ is dictated by the convolution of the Gaussian Imry-Wortis disorder landscape with the kinetic arrest boundary:
\begin{equation}
    X(T) = \int_{-\infty}^{\infty} \Theta(T - T_c(\mathbf{r})) P(T_c) \, d T_c,
\end{equation}
subject to the dynamic arrest constraint $\tau(T, \mathbf{E}) \le t_{\mathrm{exp}}$. On cooling, as temperature falls below $T_c$, self-generated elastic clamping elevates the local activation barrier $U_{\mathrm{eff}} = \Delta f^{\ddagger} + U_{\mathrm{el}}$. When $T \to T_K$, the boundary velocity $v \propto \Gamma(-\delta F/\delta \phi) \to 0$, locking the system into an arrested state with a non-zero residual metallic fraction $X_{\mathrm{stall}} \approx 0.60$. As shown in Figure 4, this mechanism prevents the exponential divergency characteristic of bulk single crystals, stabilizing a low-temperature resistivity plateau of $\rho \approx 4.05 \times 10^{-3}~\Omega\cdot\mathrm{cm}$ below $50$~K. 

Upon warming from the kinetically arrested state, the system exhibits delayed de-arrest due to the persistent elastic clamping barrier. Initial thermal activation allows localized stress relaxation, which manifests as a broad peak in resistivity centered at $T \approx 55$~K ($\rho \approx 4.5 \times 10^{-3}~\Omega\cdot\mathrm{cm}$) before extensive phase boundary depinning restores the high-temperature metallic phase near $185$~K. The calculated $\rho(T)$ profile matches qualitatively with the experimental temperature dependence of resistivity and associated thermal hysteresis in epitaxially grown $\mathrm{V_2O_3}$ thin films on sapphire substrate \cite{de}. It demonstrates that kinetic arrest and substrate clamping provide a plausible description of thermal metastability in strained $V_{2}O_{3}$ thin films without requiring an applied electric field. It may, however, be noted that this is at best a qualitative representation of the experimental results, sinnce the experimental inputs used in the simulation are approximate in nature. Also, we have not included the effect of electrothermal Joule heating, which will be discussed in the next subsection.

\subsection{Electrothermal Joule Heating versus Non-Thermal Field-Driven De-Arrest}
A comprehensive assessment of resistive switching in $\mathrm{V_2O_3}$ requires evaluating the proposed electric-field-induced barrier lowering against electrothermal effects, notably Joule self-heating. In macroscopic planar devices and two-terminal micro-gap structures under DC or long-pulse voltage excitation, a substantial body of experimental and numerical work has demonstrated that threshold switching is often dominated by electrothermal filamentation \cite{stolier2014,janod2015,valmianski2018}. In this electrothermal picture, current crowding along the most conducting percolation path initiates local Joule heating ($P = J^2 \rho$). Because the resistance drops precipitously across the insulator-to-metal transition, positive feedback causes a localized thermal runaway, elevating the local channel temperature above $T_c$ and stabilizing a conducting metallic filament.

In contrast, our continuum formulation treats the electric field coupling $-\alpha |\mathbf{E}|$ as an idealized, isothermal electronic activation mechanism analogous to dielectric softening of the lattice distortion, field-assisted carrier de-trapping, or Zener-Mott tunneling. This treatment serves as an analytical limiting case, elucidating how a pure electrostatic field modulates the kinetic arrest landscape independently of macroscopic temperature increases.

Experimentally, several diagnostic criteria can distinguish between an electrothermal mechanism and non-thermal field-driven de-arrest:
\begin{enumerate}
    \item {\it Pulse-Width and Timescale Separation:} Electrothermal filamentation is constrained by the thermal time constant of the film-substrate architecture ($\tau_{th} \approx C_v d^2 / \kappa_{th} \sim 10 - 100$ ns). Under ultrafast, sub-nanosecond electrical pulses ($\tau_{pulse} \ll \tau_{th}$), thermal accumulation is strongly suppressed, allowing intrinsic electronic barrier lowering and field-induced de-arrest to be probed directly.
    \item {\it Device Geometry and Power Dissipation:} In nanoscale gaps ($< 50$~nm), critical switching fields ($E_{th} \sim 10^5$~V/cm) can be reached at low absolute voltages ($V < 0.5$~V) and sub-microamp currents, dramatically reducing dissipated power and attenuating Joule heating.
    \item {\it Threshold Temperature Scaling ($dE_{th}/dT$):} Purely electrothermal models predict that the threshold switching power scales with the temperature difference to the transition, $P_{th} \propto (T_c - T_{ambient})$, yielding a square-root voltage dependence $V_{th} \propto \sqrt{T_c - T_{ambient}}$. In contrast, our kinetic barrier-lowering model predicts an approximately linear dependence, $E_{th}(T) \propto (U_0 + \langle U_{el} \rangle) - \beta T$ .
\end{enumerate}
Thus, while steady-state switching in macroscopic $\mathrm{V_2O_3}$ devices typically involves electrothermal feedback, our theory provides the requisite framework for understanding the non-thermal limit where field-modulated barrier collapse initiates domain de-pinning.

\subsection{Broader Context: Kinetic Arrest in Disordered Systems}
The competition between quenched disorder and long-range constraints identified in $\mathrm{V_2O_3}$ represents a general paradigm across disordered condensed matter systems undergoing sharp phase transformations. A relevant parallel exists in magnetic and multiferroic systems \cite{sbroy2013,chaddah2018}, and highly disordered superconducting thin films, such as amorphous NbGe (a-NbGe), MoGe, and granular NbN, utilized in Superconducting Nanowire Single-Photon Detectors (SNSPDs) and studies of the superconductor-insulator transition \cite{{natarajan2012}}. In the last classses of materials, structural amorphization induces local fluctuations in the superconducting pairing potential, generating a distribution of local critical temperatures $\delta T_c(\mathbf{r})$ analogous to the Imry-Wortis landscape treated here. Under transport current bias or non-equilibrium optical excitation, the breakdown of superconductivity proceeds via the nucleation, pinning, and thermal relaxation of normal-state hotspots. In both $\mathrm{V_2O_3}$ and disordered superconducting films, the system settles into pinned, non-equilibrium configurations, and external fields drive avalanche-like de-arrest dynamics, highlighting common kinetic principles across disparate quantum materials.

\section{Conclusions}
In conclusion, we have formulated a continuum Time-Dependent Ginzburg-Landau framework that models the non-equilibrium transformation kinetics of the first-order metal-insulator transition in epitaxial $\mathrm{V_2O_3}$ thin films. By integrating an Imry-Wortis disorder landscape with non-local elastic interactions under thin-film clamping conditions, the theory captures the microscopic origin of phase boundary pinning, slow relaxation, and persistent phase coexistence.

Our dynamic derivation of the kinetic arrest condition $T_K(\mathbf{E})$ clarifies how the divergence of internal relaxation times relative to experimental timescales freezes the system into a metastable, two-phase configuration. The simulated transformation kinetics exhibit an Avrami exponent suppression ($n < 1$), providing a quantitative signature of strain-hindered domain growth. Furthermore, the reduction of activation barriers by an external electric field provides a consistent physical picture for threshold de-arrest and multi-level resistance states in the non-thermal limit, complementing conventional electrothermal switching models. By connecting microscopic domain interactions to macroscopic transport observables, this framework offers a useful basis for interpreting non-equilibrium phase transformations in correlated functional oxides and related disordered systems.

\ack
S.B.R. acknowledges support from Ramakrishna Mission Vivekananda Educational and Research Institute (RKMVERI), Belur Math, Department of Physics, Indian Institute of Technology Bombay, and UGC-DAE Consortium for Scientific Research, Indore. Useful discussions with several academic colleagues are gratefully acknowledged. Some Help of AI has been used in this manuscript, particularly in the numerical simulation in section 4.4 with cloud computing but with effective crosschecking .

\clearpage
\end{document}